\documentclass[]{pasj01}

\Received{2023/08/09}
\Accepted{2023/12/27}
 
\usepackage[switch,mathlines]{lineno}
 
\begin{document} 

\title{ 
Trigonometric parallax and proper motion of Sagittarius A* measured by VERA using the new broad-band back-end system OCTAVE-DAS}

\author{Tomoaki \textsc{Oyama}\altaffilmark{1,*}%
}


\altaffiltext{1}{Mizusawa VLBI Observatory, National Astronomical Observatory of Japan, 2-12 Hoshiga-oka, Mizusawa, Oshu-shi, Iwate 023-0861, Japan}

\author{Takumi \textsc{Nagayama}\altaffilmark{1}%
}

\author{Aya \textsc{Yamauchi}\altaffilmark{1}%
}

\author{Daisuke \textsc{Sakai}\altaffilmark{1}%
}

\author{Hiroshi \textsc{Imai}\altaffilmark{2,3}%
}

\author{Mareki \textsc{Honma}\altaffilmark{1,4}%
}

\author{Yu \textsc{Asakura}\altaffilmark{1}%
}

\author{Kazuhiro \textsc{Hada}\altaffilmark{1,5}%
}

\author{Yoshiaki \textsc{Hagiwara}\altaffilmark{6}%
}

\author{Tomoya \textsc{Hirota}\altaffilmark{1,5}%
}

\author{Takaaki \textsc{Jike}\altaffilmark{1,5}%
}

\author{Yusuke \textsc{Kono}\altaffilmark{5,7}%
}

\author{Syunsaku \textsc{Suzuki}\altaffilmark{7}%
}

\author{Hideyuki \textsc{Kobayashi}\altaffilmark{5,7}%
}

\author{Noriyuki \textsc{Kawaguchi}\altaffilmark{7}%
}


\altaffiltext{2}{Amanogawa Galaxy Astronomy Research Center, Graduate School of Science and Engineering, Kagoshima University, 1-21-35 Korimoto, Kagoshima 890-0065, Japan}

\altaffiltext{3}{Center for General Education, Institute for Comprehensive Education,
Kagoshima University, 1-21-30 Korimoto, Kagoshima 8900-0065, Japan}

\altaffiltext{4}{Department of Astronomy, School of Science, The University of Tokyo, 7-3-1 Hongo, Bunkyo-ku, Tokyo
113-0033, Japan}

\altaffiltext{5}{The Graduate University for Advanced Studies, SOKENDAI, 
2-21-1 Osawa, Mitaka, Tokyo 181-8588, Japan}

\altaffiltext{6}{Toyo University, 5-28-20 Hakusan, Bunkyo-ku, Tokyo 112-8606, Japan}

\altaffiltext{7}{Mizusawa VLBI Observatory, National Astronomical Observatory of Japan, 2-21-1 Osawa, Mitaka, Tokyo 181-8588, Japan}


\email{t.oyama@nao.ac.jp}




\maketitle

\begin{abstract}
We successfully measured the trigonometric parallax of Sagittarius A* (Sgr A*)
to be $117\pm17$ micro-arcseconds ($\mu$as) using the 
VLBI Exploration of Radio Astrometry (VERA) with the newly developed broad-band signal-processing system named ``OCTAVE-DAS.''
The measured parallax corresponds to a Galactocentric distance at the Sun of $R_0 = 8.5^{+1.5}_{-1.1}$ kpc.
By combining the astrometric results with VERA and the Very Long Baseline Array (VLBA) 
over a monitoring period of 25 years,
the proper motion of Sgr A* is obtained to be 
$(\mu_\alpha, \mu_\delta) = (-3.133\pm0.003, -5.575\pm0.005)$ mas yr$^{-1}$
in equatorial coordinates, corresponding to
$(\mu_l, \mu_b) = (-6.391\pm0.005, -0.230\pm0.004)$ mas yr$^{-1}$
in Galactic coordinates.
This gives an angular orbital velocity of the Sun 
of $\Omega_\odot = 30.30 \pm 0.02$ km s$^{-1}$ kpc$^{-1}$.
We find upper limits to the core wander, $\Delta \theta < 0.20$ mas (1.6 AU), peculiar motion, $\Delta \mu < 0.10$ mas yr$^{-1}$ (3.7 km s$^{-1}$), and acceleration, $a < 2.6$ $\mu$as yr$^{-2}$ (0.10 km s$^{-1}$ yr$^{-1}$) for Sgr A*.   
Thus, we obtained upper mass limits of $\approx$ 3 $\times$ 10$^{4}$$M_{\odot}$ and  $\approx$ 3 $\times$ 10$^{3}$$M_{\odot}$ for the supposed intermediate-mass black holes at 0.1 and 0.01 pc from the Galactic center, respectively.
\end{abstract}



\section{Introduction}
The distance from the local standard of rest (LSR) 
to the Galactic center, $R_0$, 
and the angular velocity of the Sun, 
$\Omega_\odot = (\Theta_0+V_\odot)/R_0$,
which is the angular velocity of the Galaxy as the sum of the Galactic rotation at the LSR ($\Theta_0$) 
and the solar motion in the Galactic rotation direction ($V_\odot$), 
are fundamental parameters for providing the size, rotation velocity, and hence, mass of the Galaxy.
Recently, VLBI astrometry of the Galactic star-forming regions 
hosting strong maser emission estimated these parameters at relative accuracy levels of a few percent, 
being reported to be $R_0=8.15\pm0.15$ kpc and 
$\Omega_\odot = 30.32\pm0.27$ km s$^{-1}$ kpc$^{-1}$
by Bar and Spiral Structure Legacy (BeSSeL) \citep{rei19} and 
$R_0=7.92\pm0.16({\rm stat.})\pm 0.3({\rm sys.})$ kpc and
$\Omega_\odot = 30.17\pm0.27({\rm stat.})\pm 0.3({\rm sys.})$ km s$^{-1}$ kpc$^{-1}$ 
from the VLBI Exploration of Radio Astrometry (VERA) \citep{vera20}.
Here, stat. and sys. mean the statistical and systematic errors, respectively.
Comparison of the results of BeSSeL and VERA shows that  
$\Omega_\odot$ is consistent within mutual error ranges 
and at a relative error of $\approx 1$\%.
The geometric distance estimated using the star orbit around Sgr A*, which is a supermassive black hole (SMBH) (\cite{eht22}), located at the Galactic dynamical center, 
are $R_0=8.178\pm0.013({\rm stat.})\pm0.022({\rm sys.})$ kpc \citep{gra19}, $R_0=8.275\pm0.009({\rm stat.})\pm0.033({\rm sys.})$ kpc \citep{gra21}, and
$R_0=7.971\pm0.059({\rm stat.})\pm0.032({\rm sys.})$ kpc \citep{do19}.
Comparing the five estimations of $R_0$, 
reveals a significant discrepancy of $\approx 0.4$ kpc, 
corresponding to $\approx 4$\% uncertainty.
Even though this value will be reduced in the future,  
independent ways of measurement would be required.

The most direct way to determine the fundamental parameters is to measure
the trigonometric parallax and the proper motion of Sgr A*.
The proper motion measurement has been conducted with the Very Large Array (VLA) and the Very Long Baseline Array (VLBA) (\cite{bac99,rei99,rei04,rei20,xu22}).
The angular velocity has been derived to be 
$\Omega_\odot = 30.39\pm0.04$ km s$^{-1}$ kpc$^{-1}$ \citep{rei20},
which is consistent with that derived from the Galactic maser astrometry data (\cite{rei19,vera20}).
However, the parallax measurement for Sgr A* remains an issue for the VLBI astrometry. 
Measurements of the distance to the Galactic center using the VLBI astrometry technique
were conducted through observations of Sgr B2 in the vicinity of Sgr A*.
The estimates of $R_0$ were $7.1\pm1.5$ kpc from statistical parallax  (\cite{rei88}), $7.9^{+0.8}_{-0.7}$ and $7.7^{+3.0}_{-1.7}$ kpc from trigonometric parallax (\cite{rei09}; \cite{sak23}).
Although these observations and results are pioneering and consistent with the results obtained using other methods, ultimately 
it is crucial to directly measure the trigonometric parallax of Sgr A*.

We aimed to measure the parallax of Sgr A* using VERA. 
VERA consists of four 20-m antennas with a dual-beam receiver system for 22 and 43 GHz, enabling us to simultaneously observe an adjacent phase calibrator (positional reference) (e.g., an extra-Galactic continuum source) and a target (e.g., a Galactic maser source)  within \timeform{2.2D} \citep{kaw00}.
These antennas are located at Mizusawa (hereafter MIZ), Iriki (IRK), 
Ogasawara (OGA), and Ishigaki-jima (ISG) in Japan
and the six baselines range from a minimum of 1200 km 
to a maximum of 2300 km \citep{kob03}, 
corresponding to a synthesized beam size of 0.5 mas at a radio frequency of 43 GHz. 
In the case of the best parallax precision of 10 micro-arcseconds ($\mu$as) for VERA \citep{nag20},
it is expected that the Sgr A*'s parallax of 1/(8 kpc) = 125 $\mu$as can be
measured at a relative error of (10 $\mu$as)/(125 $\mu$as) = 8\%.

\begin{figure}
\begin{center}
\includegraphics[width=8cm]{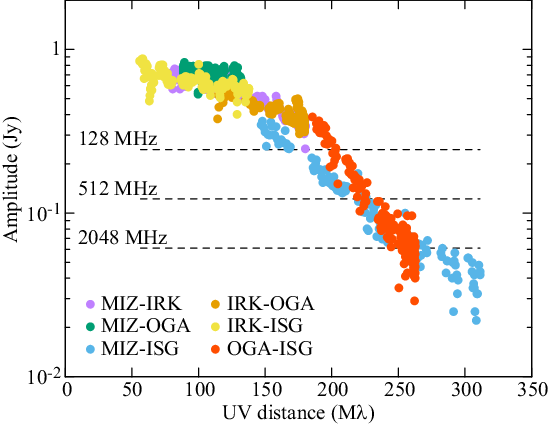} 
\end{center}
\caption{Visibility amplitude vs $UV$ distance plot of Sgr A* for the VERA 43-GHz observation. 
The visibility data are shown after self-calibration.
The dashed lines show 5$\sigma$ detection limits of bandwidths of 128, 512, and 2048 MHz,
where $\sigma$ is the baseline-based noise level estimated from 
the system noise temperature of 300 K, the averaging time is 60 s, 
the antenna dish diameter is 20 m, and the antenna aperture efficiency is 0.5. (Color online)}
\label{fig:1}
\end{figure}

However, the issue for VERA observations of Sgr A* was array sensitivity.
The VLBI fringe of Sgr A* could not be detected at relatively long baselines of VERA
using the previous signal-processing system yielding a bandwidth of 128 MHz per beam.
Figure \ref{fig:1} shows the amplitude vs $UV$ distance plot of Sgr A* 
for VERA 43-GHz observations with the new signal-processing system (section 2).
The visibility amplitude decreases with longer baselines because of 
the angular broadening by the diffractive interstellar scattering \citep{lo98,gwi14,joh18}.
In the case of the bandwidth of 128 MHz, 
the system noise temperature of 300 K, the averaging time of 60 s,
the antenna dish diameter of 20 m, and the aperture efficiency of 0.5,
the baseline-based noise level is estimated to be $\sigma \approx 48$ mJy.
The fringe detection over $5\sigma$ 
is limited to the short baselines within 180 M$\lambda$ (1300 km).
However, if the bandwidth expands to 2048 MHz,
the detectable baseline length extends up to 250 M$\lambda$ (1800 km)
which almost covers the entire range of the VERA baselines.

VERA has been regularly operated for about 18 yr since 2005. 
We have started astrometric observations of scientifically important Galactic maser sources associated with strong reference calibrators \citep{vera20}.
However, there still remains numerous maser sources that have not yet been observed owing to a lack of detectable calibrators using the current observing system with insufficient sensitivity. Therefore, a sensitivity upgrade for reference continuum sources is crucial to increase the number of target sources for VERA Galactic maser astrometry. 
A previous VERA calibrator survey \citep{pet07} showed that 533 out of $\sim$2500 continuum sources were detected  within \timeform{6D} of the Galactic plane, within $11^{\circ}$ of the Galactic center, or within \timeform{2.2D} of SiO or H$_2$O masers.  If we assume a uniform number density of calibrators in the sky plane, then $\sim$2700 calibrators are required in the entire sky to be detected within \timeform{2.2D} around any of the Galactic maser sources. Therefore, the survey yields a sufficiency rate of $\sim$$20\%$. If the bandwidth is a factor of 8 wider, 
it is possible to detect a sufficient number of calibrators within \timeform{2.2D} of any sky position. 
Therefore, we developed a new 
broad-band digital back-end and software correlator system, that provides a bandwidth of 2048 MHz for Sgr A*.
Because of fringe detections at long baselines,
the accuracy of the position measurement
is expected to be improved by a factor of (1800 km)/(1300 km) = 1.4.

In this paper, we present the new broad-band digital back-end and software correlator 
system installed at the VERA and Mizusawa Correlation Center 
and report the first astrometric results, using these systems, yielding the detection of the trigonometric parallax of Sgr A*.
In section 2, the configuration and specifications of the proposed broad-band system are presented. 
Section 3 presents the observations and data analyses.
In section 4, we verify the system and report the astrometric results.
In section 5, we discuss the astrometric position error and dynamics of Sgr A* 
through a comparison with previous observations. 

\section{Instrument}

\subsection{OCTAVE-family}

VERA's new data acquisition and correlator system comprises Optically Connected Array for VLBI Exploration (OCTAVE) instruments \citep{oya12,oya16}, and 
has been developed since the previous OCTAVE experiments 
\citep{kaw01,fuj01,has04,tak08,doi09,kon12}. 
The OCTAVE data acquisition system (OCTAVE-DAS) has been developed on the basis of the VLBI Standard Hardware Interface (VSI-H)\footnote{http://www.vlbi.org/vsi/index.html.} and the VLBI Data Interchange Format (VDIF)\footnote{http://www.vlbi.org/vdif/index.html.} specifications.
It consists of high-speed samplers called OCTAD (OCTAve A/D converter), media converters between VSI-H and VDIF called OCTAVIA (OCTAve VSI Adapter), high-speed recorders called VSREC (VDIF Software RECorder), OCTADISK and OCTADISK2 (OCTAve DISK drive), and field programmable gate array-based hardware and software correlators called OCTACOR and OCTACOR2 (OCTAve CORrelator). 

OCTAD digitizes radio frequency (up to 26 GHz) broad-band signals with eight quantization (3-bit) levels at a sampling rate of 8192 MHz up
to 20480 MHz. OCTAD has a digital baseband converter (DBBC) function, in which the number of output baseband streams from 4 to 32 
and a bandwidth per stream can be selected from 8 to 2048 MHz discretely ($2^n$ MHz, where $n=3$--11).
OCTAVIA uses the User Datagram Protocol (UDP) for data transfer and the Transmission Control Protocol (TCP) for device control.

OCTADISK is a disk recorder implemented on a field programmable gate array.
VSREC is one of our VDIF software libraries, but here it refers to a disk recorder system that uses this software. This recorder system consists of commercial off-the-shelf (COTS) servers and raid-boxes. 
OCTADISK2 is a successor to OCTADISK and is used in conjunction with OCTAD.
OCTADISK2 was developed by Elecs Industry as an upgraded version of the VSREC system and is based on its experience in developing the VSREC system. OCTADISK2 can not only record (at a maximum rate of 32 Gbps) but can also playback functions (at a maximum rate of 16 Gbps).

OCTACOR2 is equipped with an FX-type software correlator based on the gico3 (main correlation software; \cite{kim02}), softcos (pre-, post-correlation software) and  Flexible Image Transport System (FITS) file generation software. This OCTACOR2 system was previously installed at the NAOJ Mitaka campus. After it was moved to the NAOJ Mizusawa campus in April 2015, it was renamed the Mizusawa software correlator system.
For more detailed specifications and technical information of these instruments belonging to the OCTAVE family, see \citet{oya12} and forthcoming technical papers.

\subsection{New broad-band observing system for VERA}

\begin{figure*}
 \begin{center}
  \includegraphics[width=16cm]{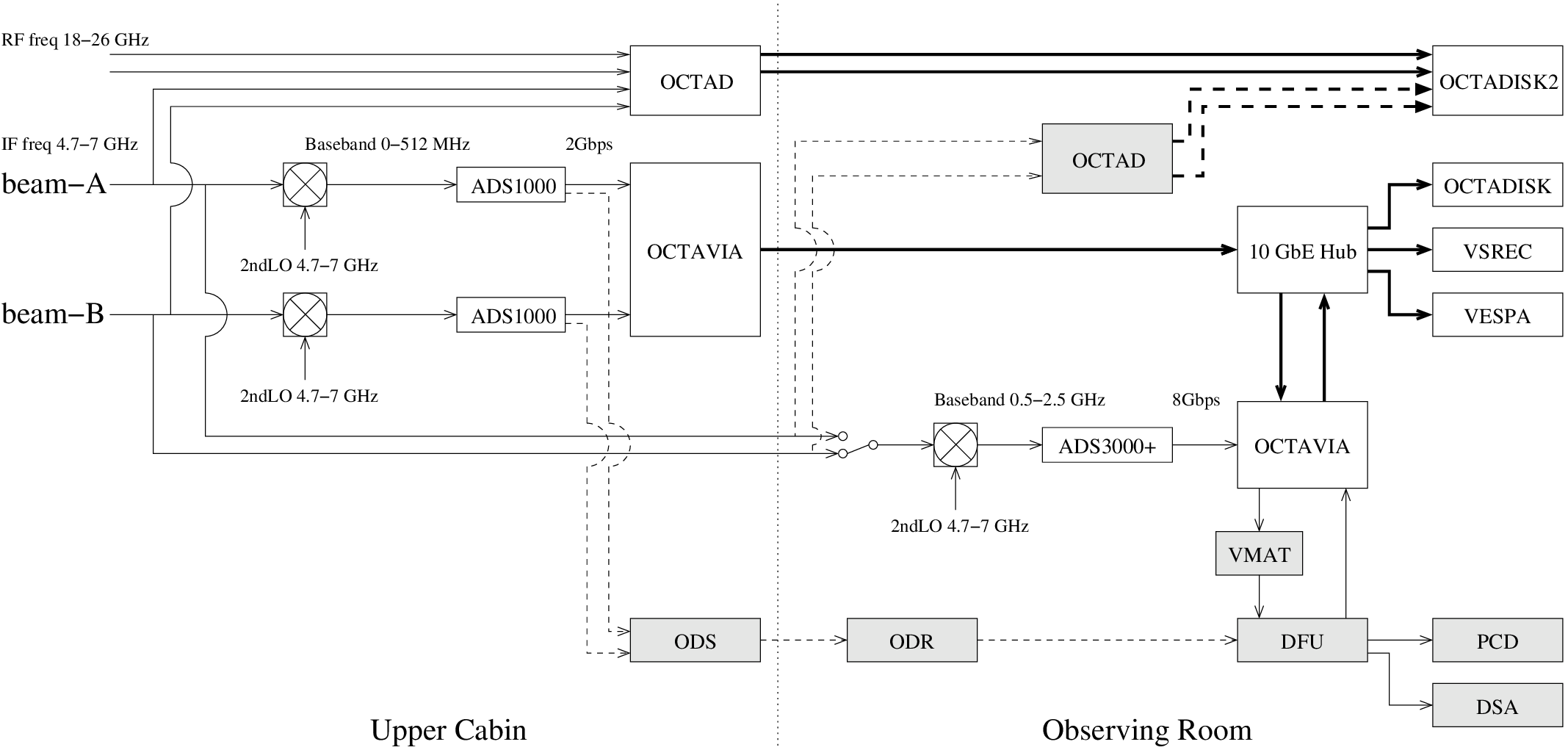}
 \end{center}
\caption{Block diagram of the present VERA back-end system for broad-band observations. 
Instruments shown in gray rectangles are those included in the conventional system. 
Thin arrows indicate the flows of electrical and digitized signals; OGA stations use the conventional signal lines shown by broken arrows via ODS and ODR.
Thick arrows show the signal paths in the 10-Gbps Ethernet network. 
OCTAD is located in the observation room (MIZ and IRK, until summer 2021) or in the upper cabin of the telescope.
OCTAVIA, OCTADISK, VSREC, OCTAD and OCTADISK2 belong to the OCTAVE system.}
\label{fig:2}
\end{figure*}

Figure \ref{fig:2} shows a block diagram of the VERA back-end system 
 with the OCTAVE-DAS for broad-band observations using a dual-beam receiver system.
 
VERA has two main types of broad-band observing systems. 
The conventional type uses ADS1000 and ADS3000+ formatters developed by NICT (\cite{tak06}; \cite{tak10}). The other uses OCTAD.
The right and left-hand circularly polarized (RCP and LCP) signals received by the two receivers, called beam-A and beam-B receivers. The both LCP signals are converted to signals of an intermediate frequency (IF) band of 4.7--7 GHz. The receiver system (beam-A only) was upgraded to conduct broad-band (16 Gbps) dual-polarized observations at K and Q bands from August 2019 for all stations: subsequently, demonstrative performance evaluation and scientific observations were conducted, which yielded reasonable and valuable results \citep{hag22, tak23}.

The IF signals are split and converted further down to the baseband streams to be sampled using the ADS1000 and ADS3000+ formatters, both at four quantized levels. The two types of formatters outputs have each one and four baseband streams, each with a bandwidth of 512 MHz.
The signals from the two receivers are input into two ADS1000s.
Moreover, either beam-A or beam-B signal selected by an analog switch is input to the ADS3000+ formatter.
The basic observing mode is conducted at a recording rate of 4 Gbps using only ADS1000s. 
The boost mode (12 Gbps), which uses the ADS3000+ in addition to the ADS1000s, has 
been in use since 2013.

In the conventional VERA system, signal output from ADS1000 is transferred to the observing room via the optical digital transmitter (ODS) and the optical digital receiver (ODR); ODS and ODR use 
 the Asynchronous Transfer Mode (ATM) protocol.  
ODS and ODR are still used only at OGA station, but will be replaced by OCTAVIA at all stations in the near future.
The transferred signals are filtered by DFU (the digital filter unit; \cite{igu05}) and then input to a phase calibration detector (PCD) and digital spectrum analyzer (DSA), which were also input to a tape recorder (DIR2000) via a VSI-interface at a rate of 1 Gbps; however, DIR2000 has already been decommissioned.
In the present system, signals compliant with the VSI-H output from the DFU, ADS1000, and ADS3000+ are converted to signals compliant with the VDIF (10 GbE) format in OCTAVIA and transferred to recorders (OCTADISK and VSREC) and PolariS (the software polarization spectrometer; \cite{kam14}, \cite{miz14}). 
To monitor and calibrate the relative instrumental delays between the signal paths through the dual-beams \citep{hon08a}, the signals from two ADS1000s are transferred to a PCD via OCTAVIA, a VLBI Multi Adapter and Tester (VMAT), and DFU.
OCTAVIA and VMAT are connected using two MDR-80 pin connector cables (VSI-H, 64-MHz-clock 32 bits).
VMAT converts low voltage differencial signaling (LVDS) compliant with VSI-H signals to emitter coupled logic (ECL) level signals and outputs them to four MDR-68 pin connector cables (32-MHz-clock 32 bits).
If we want to make a conventional observation using VERA, the combination array of the Korean VLBI Network (KVN) and VERA, namely KaVA and the East Asia VLBI Network (EAVN; \cite{an18}) at a common, lower recording rate, e.g., 1 Gbps, we can transmit such signals filtered by the DFU to OCTADISK via OCTAVIA.
When we use a 4 or 12-Gbps recording mode, we use VSREC as the recorder.

The IF signals of 4.7--7 GHz are branched and also input to the OCTAD.
There are OCTADs at the OGA and ISG stations located in the upper cabin.
Those at the other stations were installed in the operations room by the summer of 2021, but they have been moved to the upper cabin of the telescope at present.
The OCTAD installed in the upper cabin can accept the RCP and LCP RF signals of 18--26 GHz
from K-band receivers as well as the IF signals of 5--7 GHz, and 
 not only samples IF signals but also converts 4096-MHz data to four set of 512-MHz data.
The OCTAD and OCTADISK2 are directly connected without a 10GbE hub.
This setup is capable of recording in a 16-Gbps mode (512-MHz bandwidth $\times$ 8 streams). The detailed specifications are listed in table \ref{tab:1}.

\begin{table*}
\tbl{Observing modes for Sgr A* with OCTAVE-DAS}{
\begin{tabular}{cccccc}
\hline
\multicolumn{1}{c}{Mode}           &
\multicolumn{1}{c}{Input beams}  &
\multicolumn{1}{c}{Input bandwidth}  &
\multicolumn{1}{c}{Input bits/sample}   &
\multicolumn{1}{c}{}Output bandwidth        &       
\multicolumn{1}{c}{Output streams}           \\      
\multicolumn{1}{c}{}            &
\multicolumn{1}{c}{}            &
\multicolumn{1}{c}{(MHz)$\times$streams}            &
\multicolumn{1}{c}{}            &
\multicolumn{1}{c}{(MHz)}            &
\multicolumn{1}{c}{}               \\  
\hline
ADS1000$^{*}$  & A\&B & 512 & 2 & 512 & 2 \\
\hline
ADS1000$^{\dagger}$  & A\&B & 512 & 2 & 512 & 2 \\
ADS3000+$^{\dagger}$ & B & 512$\times$4 & 2 & 512 & 4 \\
\hline
OCTAD  & A\&B & 4096  & 3 & 512 & 8  \\ 
\hline
\end{tabular}}
\label{tab:1}
\begin{tabnote}
$^{*}$ Basic mode. \\
$^{\dagger}$ Boost-up mode.
\end{tabnote}
\end{table*}

\subsection{Data correlation for broad-band observations with VERA}

The correlations were processed using the software FX-type correlator, OCTACOR2.
In 12-Gbps observations, the following three types of data are recorded simultaneously: (1) one stream of beam-A data sampled by ADS1000 (St1), (2) one stream of beam-B data sampled by ADS1000 (St2), and (3) four streams of beam-B data sampled by ADS3000+ (St3 to St6).
St1, St2 and St4 record the same frequency band.
OCTACOR2-PCs correlate data from VERA stations for each stream from St1 to St6.
In addition, St2 and St4 data of each station are correlated to calibrate the delay difference resulting from the different paths after splitting.
Because the delay difference between St1 and St2 can be calibrated using the dual-beam delay calibration system \citep{hon08a}, we can perform phase calibration between St1 and St3--St6.

In the case of the 16-Gbps observations, the following two types of data are recorded simultaneously: (1) four streams of beam-A data sampled by OCTAD (St7 to St10) and (2) four streams of beam-B data sampled by OCTAD (St11 to St14).
The four streams from both beams correspond to each other (e.g., St7 and St11).
All data from St7 to St14 are also correlated using OCTACOR2. 
In addition, St8 and St12 data of each station are correlated to calibrate 
the relative instrumental delays between the signal paths through the dual-beams in the same manner as in the calibration between St1 and St2.

\section{Observations and Data analysis}

\subsection{Observations}

The VLBI observations of Sgr A* using VERA 
were conducted over 26 epochs between 2014 and 2020.
The observation dates are listed in table \ref{tab:2}.
The epoch spacing was approximately 1 year until the 6th epoch (until 2017).
Afterward, we observed more densely at 1--2 month cadence 
to measure the parallax.

\begin{table*}
\tbl{Position offsets of J1745$-$2820 relative to Sgr A*.}{
\begin{tabular}{rccrrrr}
\hline
\multicolumn{1}{c}{No}    &
\multicolumn{1}{c}{Obs.}  &
\multicolumn{1}{c}{Date}  &
\multicolumn{1}{c}{$\Delta \alpha$} &
\multicolumn{1}{c}{$\Delta \delta$} &
\multicolumn{1}{c}{$\Delta \alpha$} &
\multicolumn{1}{c}{$\Delta \delta$} \\
\multicolumn{1}{c}{} &
\multicolumn{1}{c}{code}   &
\multicolumn{1}{c}{(year)} &
\multicolumn{1}{c}{(mas)}  &
\multicolumn{1}{c}{(mas)}  &
\multicolumn{1}{c}{(mas)}  &
\multicolumn{1}{c}{(mas)}  \\
\hline
1  & r14089b & 2014.244 & $0.008\pm0.062$ & $0.070\pm0.188$ & $56.948\pm0.151$ & $100.460\pm0.205$ \\
2  & r14113b & 2014.310 & $0.146\pm0.062$ & $0.030\pm0.188$ & $57.086\pm0.151$ & $100.560\pm0.205$ \\
3  & r14137a & 2014.375 & $0.458\pm0.063$ & $0.568\pm0.189$ & $57.398\pm0.152$ & $101.098\pm0.207$ \\
4  & r15085b & 2015.233 & $2.913\pm0.071$ & $5.016\pm0.198$ & $59.853\pm0.155$ & $105.546\pm0.214$ \\
5  & r16092b & 2016.251 & $6.213\pm0.063$ &$10.957\pm0.188$ & $63.153\pm0.152$ & $111.487\pm0.205$ \\
6  & r17085b & 2017.233 & $9.295\pm0.062$ &$16.295\pm0.188$ & $66.235\pm0.151$ & $116.825\pm0.205$ \\
7  & r17360b & 2017.986 &$11.670\pm0.059$ &$20.779\pm0.186$ & $68.610\pm0.150$ & $121.309\pm0.203$ \\
8  & r18032b & 2018.088 &$11.844\pm0.060$ &$21.053\pm0.186$ & $68.784\pm0.150$ & $121.583\pm0.203$ \\
9  & r18092a & 2018.252 &$12.491\pm0.059$ &$21.787\pm0.185$ & $69.431\pm0.150$ & $122.317\pm0.203$ \\
10 & r18135a & 2018.370 &$12.974\pm0.060$ &$22.775\pm0.187$ & $69.914\pm0.151$ & $123.305\pm0.204$ \\
11 & r18275a & 2018.753 &$14.278\pm0.065$ &$24.651\pm0.193$ & $71.218\pm0.152$ & $125.181\pm0.209$ \\
12 & r18306a & 2018.838 &$14.499\pm0.063$ &$25.081\pm0.189$ & $71.439\pm0.152$ & $125.611\pm0.206$ \\
13 & r18312a & 2018.855 &$14.552\pm0.063$ &$25.553\pm0.190$ & $71.492\pm0.152$ & $126.083\pm0.207$ \\
14 & r18358c & 2018.981 &$14.785\pm0.062$ &$25.588\pm0.187$ & $71.725\pm0.151$ & $126.118\pm0.204$ \\
15 & r19008b & 2019.022 &$14.919\pm0.062$ &$26.126\pm0.188$ & $71.859\pm0.151$ & $126.656\pm0.205$ \\
16 & r19037b & 2019.101 &$15.134\pm0.060$ &$26.822\pm0.186$ & $72.074\pm0.150$ & $127.352\pm0.203$ \\
17 & r19071b & 2019.195 &$15.431\pm0.059$ &$27.150\pm0.185$ & $72.371\pm0.150$ & $127.680\pm0.203$ \\
18 & r19113b & 2019.310 &$15.779\pm0.069$ &$27.807\pm0.198$ & $72.719\pm0.154$ & $128.337\pm0.215$ \\
19 & r19255a & 2019.699 &$17.297\pm0.065$ &$29.970\pm0.194$ & $74.237\pm0.153$ & $130.500\pm0.211$ \\
20 & r19316a & 2019.866 &$17.731\pm0.062$ &$31.503\pm0.191$ & $74.671\pm0.151$ & $132.033\pm0.208$ \\
21 & r20006b & 2020.016 &$18.031\pm0.061$ &$31.644\pm0.188$ & $74.971\pm0.151$ & $132.174\pm0.205$ \\
22 & r20047b & 2020.129 &$18.470\pm0.061$ &$32.320\pm0.187$ & $75.410\pm0.151$ & $132.850\pm0.204$ \\
23 & r20051c & 2020.140 &$18.437\pm0.059$ &$32.430\pm0.186$ & $75.377\pm0.150$ & $132.960\pm0.204$ \\
24 & r20066a & 2020.181 &$18.556\pm0.060$ &$32.745\pm0.187$ & $75.496\pm0.150$ & $133.275\pm0.204$ \\
25 & r20089b & 2020.244 &$18.742\pm0.060$ &$33.163\pm0.187$ & $75.682\pm0.151$ & $133.693\pm0.204$ \\
26 & r20095d & 2020.260 &$18.869\pm0.062$ &$33.187\pm0.188$ & $75.809\pm0.151$ & $133.717\pm0.205$ \\
\hline
\end{tabular}}
\label{tab:2}
\begin{tabnote}
Column 1 lists the epoch-number IDs.
Column 2 lists the observation codes in the VERA project, and 
Column 3 lists the observation dates in years.
Columns 4 and 5 list the position offsets of J1745$-$2820 
relative to Sgr A* in RA and Dec, respectively.
These offsets were calculated using the following 
reference coordinates of the two sources: 
Sgr A* $(\alpha, \delta)_{\rm J2000}=$
(\timeform{17h45m40.035490s},\timeform{$-29$D00'28.21368"}), 
J1745$-$2820 (\timeform{17h45m52.495737s},\timeform{$-28$D20'26.28915"}).
Columns 6 and 7 are the same as Columns 4 and 5 but are  
calculated from the following reference coordinates adopted 
by \citet{rei20}:

Sgr A* (\timeform{17h45m40.0409s},\timeform{$-29$D00'28.118"}); 
J1745$-$2820 (\timeform{17h45m52.4968s},\timeform{$-28$D20'26.294"}) (section 5.2).
\end{tabnote}
\end{table*}

The target source, Sgr A*, and the background source, J1745$-$2820,
were simultaneously observed using the dual-beam system of VERA.
The separation angle between the two sources was \timeform{0.67D} (0.012 rad) at a position angle of \timeform{4D} in the equatorial reference frame. 
J1745$-$2820 has been well used as a background source in the previous VLBI astrometric observations 
toward the Galactic center direction (\cite{rei04}; \cite{rei09}; \cite{sak17}).
The bright fringe finder source, NRAO 530, was also observed 
every 80 min to calibrate the clock parameters.

Until the 6th epoch, the data from both Sgr A* and J1745$-$2820 
were taken at a frequency of 42770--43282 MHz using ADS1000. 
Afterward, the data from Sgr A* were additionally taken 
at wider frequencies of 42258--42770, 42770--43282, 43282--43794, and 43794--44306 MHz 
using ADS3000+; that is, the two systems of ADS1000 and ADS3000+ were used.
In the 23rd epoch, the data from both Sgr A* and J1745$-$2820
were additionally taken using OCTAD, with which one of the recorded frequency bands covers the same frequency band as that obtained using the ADS3000+. In other words, the three systems (ADS1000, ADS3000+, and OCTAD) were used simultaneously.
Correlation processing of all data was performed using OCTACOR2.
The frequency channel spacing and accumulation period were set as 1 MHz and 1 s, respectively.

\subsection{Data analysis}

Data reduction was performed using the Astronomical Image Processing System (AIPS).
We used Sgr A* as the phase-reference source, 
because it is stronger than the background source J1745$-$2820.
Under good weather conditions, 
Sgr A* was detected by global fringe fitting 
with a station-based signal-to-noise ratio (SNR) of over 7
at an averaging time of 60 s for 512-MHz bandwidth data and 30 s for 2048-MHz bandwidth data.
The solutions of the global fringe search and self-calibration 
obtained with the Sgr A* data were applied to those of J1745$-$2820.
The visibility amplitude was calibrated using information on system noise temperatures 
measured during the observations and the antenna gains that were determined before the observations on the basis of measurement of the effective antenna apertures. 
The amplitude and phase bandpass characteristics in each baseband stream 
were calibrated using data from scans on the bright fringe finder source, NRAO 530.
The delay differences among four IF bands in ADS3000+ (also in OCTAD)  
were also calibrated using the same scans.
The instrumental delay difference between the dual-beams was monitored and calibrated 
using artificial noise source signals injected into the dual-beam receivers that were cross-correlated \citep{hon08a}.

The tropospheric excess delay was precisely calibrated using 
the tropospheric zenith delay measured by the Global Positioning System (GPS) 
\citep{hon08b} and Niell's tropospheric mapping functions \citep{nie96}.
However, because the elevation angle of Sgr A* 
during the observations is $<$ \timeform{30D}, 
the error in the tropospheric zenith delay ($c \tau_{\rm err}$) 
leads to significant astrometric position error ($\Delta \theta_{\rm err}$).
Their relationship is theoretically expressed as 
$\Delta \theta_{\rm err} \approx c \tau_{\rm err} \cdot \Delta \sec Z / D$,
where $\Delta \sec Z$ is the difference in $\sec Z$ between 
the target and the reference sources described by the zenith angle $Z$, and
$D$ is the baseline length.
For the pair of Sgr A* and J1745$-$2820, 
the average $\Delta \sec Z$ of the four stations is 0.065.
At the beginning and end of the observations, 
$\Delta \sec Z$ increases rapidly because of the low elevation.
We flagged out the low-elevation data with $\Delta \sec Z > 0.2$ for the MIZ station 
and $\Delta \sec Z > 0.1$ for the other stations.
In the case of $\Delta \sec Z > 0.1$, 
the corresponding astrometric position error is estimated to be
$\Delta \theta_{\rm err} \approx c\tau_{\rm err} \cdot \Delta \sec Z / D > 0.2$ mas
from the tropospheric zenith delay error of 
$c \tau_{\rm err} \approx 20$ mm (\cite{hon08b}) and 
VERA's maximum baseline length of $D \approx 2300$ km.
Consequently, we must still consider the trade-off between such large position errors and a decrease in visibility data. 

The delay tracking center of J1745$-$2820 was set to
$(\alpha,\delta)_{\rm J2000} = (\timeform{17h45m52.495737s}, \timeform{-28D20'26.28915"})$
based on the VLBA calibrator list\footnote{http://www.vlba.nrao.edu/astro/calib/}. 
The errors in Right Ascension (RA) and Declination (Dec) were $\sim$7 and $\sim$9 mas, respectively.
In case of Sgr A*, we performed phase tracking again using our program at the coordinates adopted as follows: 
We set the delay tracking center of Sgr A* for each epoch
so that the position offset of J1745$-$2820 on the phase-referenced image 
($\Delta \theta_{\rm off}$) was within 1 mas.
In this case, the propagating astrometric position error is estimated to be
$\Delta \theta_{\rm off} \cdot \theta_{\rm SA} \approx$ 1 mas $\cdot 10^{-2} \approx 0.01$ mas.
Finally, the position of Sgr A* was obtained using the position offset of J1745$-$2820 
on the phase-referenced image and sources coordinate of the Sgr A* set in the delay tracking.

To investigate the spatial structure of J1745$-$2820,
we also analyzed the data observed at 22 GHz with VERA on January 28, 2019.
These data were taken at the frequencies of 21457--21969, 21969--22481, 
22481--22993, and 22993--23505 MHz using ADS3000+ (512 MHz $\times$ 4 IFs $\times$ 1 beam).
We could detect the fringe of J1745$-$2820 with an averaging time of 3 min,
although it was not detectable with the normal 1-Gbps 
recording mode.
Therefore, self-calibration imaging was performed.

\section{Results}

\subsection{Verification of the new observing system}

It is important to check the consistency of 
the astrometric results when updating the observing system.
We observed by using the three systems ADS1000, ADS3000+, and OCTAD simultaneously at the 23rd epoch.
Using the data from this epoch,
we compared the phase-referencing position of J1745$-$2820 
obtained by the three systems as listed in table \ref{tab:3}.
We confirmed that the difference among the three determined positions was only 10 $\mu$as.

\begin{figure}
\begin{center}
\includegraphics[width=8cm]{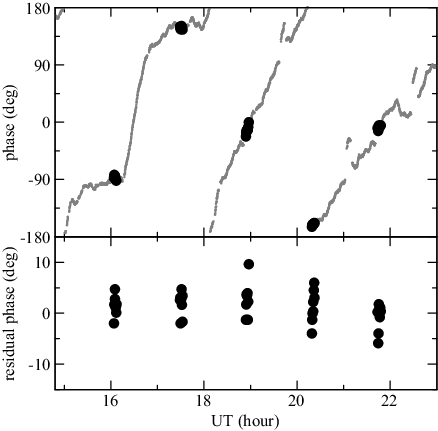} 
\end{center}
\caption{(Top) Fringe phases of the calibration data (gray dots) 
and those of NRAO 530 (black circles) obtained at the MIZ-IRK baseline at the 9th epoch.
(Bottom) Residual phase of the fringe search minus calibration data.}
\label{fig:3}
\end{figure}

Note that the length of the transmission cable from the receiver to the sampler and the phase of the local oscillator are different for ADS1000 and ADS3000+.
We need to calibrate the delay and phase differences between the two systems for phase-referencing. 
Therefore, correlation processing between the data recorded on the two systems was performed, 
and the fringe delay and phase of the common receiver output before splitting the signal through the two systems 
were monitored during the observations. 
Figure \ref{fig:3} shows a sample of
the fringe phase of the receiver noise
used for the calibration data.
We verified the data calibration using 
the fringe finder NRAO 530
which is bright enough to detect on each system.
Fringe fitting of NRAO 530 was independently performed on each system,
and the phase difference between them was obtained.
The calibration data and the fringe of NRAO 530 
are coincident well within $\sigma_\phi \approx \timeform{5D}$.
This corresponds to an astrometric position uncertainty of
$(\sigma_\phi / \timeform{360D}) \cdot \theta_{\rm beam} \approx 7$ $\mu$as 
for VERA's synthesized beam size at 43 GHz of $\theta_{\rm beam} \approx 0.5$ mas.
Therefore, this calibration method using the receiver noise 
is valid for high precision VLBI astrometry at the 10-$\mu$as level. 

\begin{figure}
\begin{center}
\includegraphics[width=8cm]{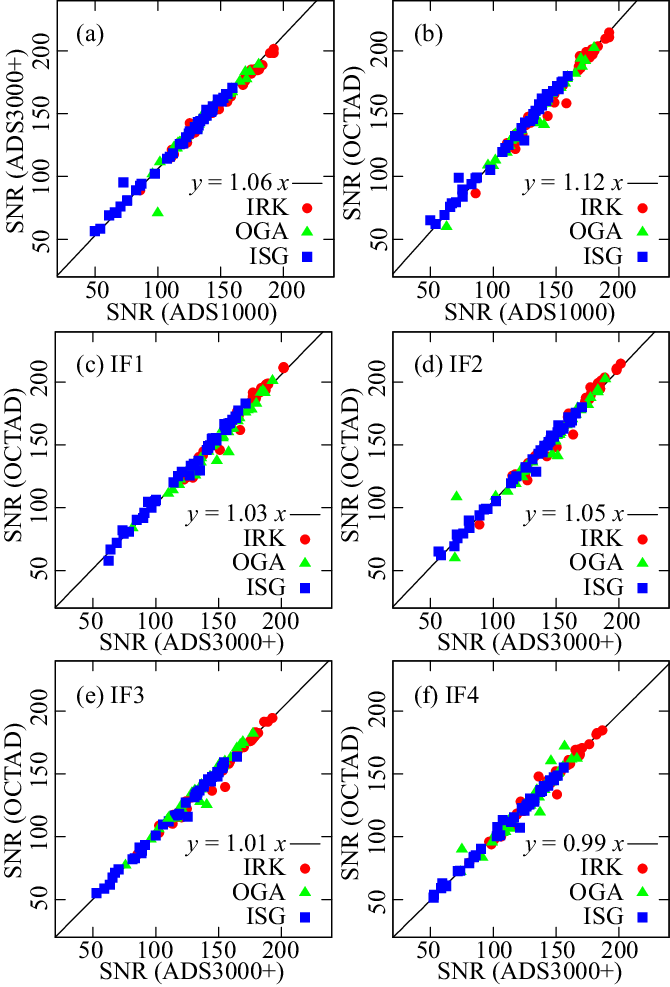} 
\end{center}
\caption{Comparison of SNRs among ADS1000, ADS3000+, and OCTAD: 
(a) ADS1000 vs ADS3000+: (b) ADS1000 vs OCTAD: 
(c)--(f) ADS3000+ vs OCTAD. 
Four IF bands (IF1, IF2, IF3, and IF4) are displayed in the individual baseband streams, each with a bandwidth of 512 MHz. 
The SNRs were obtained from the global fringe search of NRAO 530 
with an averaging time of 60 s using data from the 23rd epoch.
The reference station was set to the MIZ station; therefore the station-based SNRs
of the other three stations of IRK (red circle), OGA (green triangle), 
and ISG (blue square) are shown.
The solid line and equation at the bottom right corner of each panel
show a linear fit of $y=ax$. (Color online)}
\label{fig:4}
\end{figure}

We also checked the consistency of the SNRs among ADS1000, ADS3000+, and OCTAD.
Figure \ref{fig:4} shows the correlation plots of SNRs among the three systems.
The SNR of each system was obtained from
the global fringe fitting with an averaging time of 60 s 
using the NRAO 530 at the 23rd epoch.
The slopes of the correlation plots were obtained by linear fitting, 
and are consistent with unity at the 10\% level.
As shown in table \ref{tab:3},
the peak intensity and dynamic range of J1745$-$2820 on the phase-referenced image
were also consistent among the three systems at the 10\% level.
We conclude that the three systems are consistent at the 10-$\mu$as level 
in astrometric position and at the 10\% level in the amplitude and SNR.

\begin{table*}
\tbl{Comparison among the results with the three observing systems on 
phase-referencing position ($\Delta\alpha, \Delta\delta$),
peak intensity ($I_{\rm peak}$), and image dynamic range (DR) 
of J1745$-$2820 at the 23rd epoch.}{
\begin{tabular}{crrcc}
\hline
\multicolumn{1}{c}{Obs.}           &
\multicolumn{1}{c}{$\Delta \alpha$}  &
\multicolumn{1}{c}{$\Delta \delta$}  &
\multicolumn{1}{c}{$I_{\rm peak}$}   &
\multicolumn{1}{c}{DR}               \\
\multicolumn{1}{c}{system}           &
\multicolumn{1}{c}{(mas)}            &
\multicolumn{1}{c}{(mas)}            &
\multicolumn{1}{c}{(Jy beam$^{-1}$)} &
                                     \\
\hline
ADS1000  & $18.444\pm0.013$ & $32.426\pm0.029$ & $28.8\pm1.6$ & 18 \\ 
ADS3000+ & $18.437\pm0.013$ & $32.430\pm0.030$ & $30.8\pm1.7$ & 18 \\ 
OCTAD    & $18.439\pm0.013$ & $32.422\pm0.027$ & $33.5\pm1.6$ & 20 \\ 
\hline
\end{tabular}}
\label{tab:3}
\begin{tabnote}
\end{tabnote}
\end{table*}

\subsection{Proper motion and parallax}

\begin{figure*}
\begin{center}
\includegraphics[width=17cm]{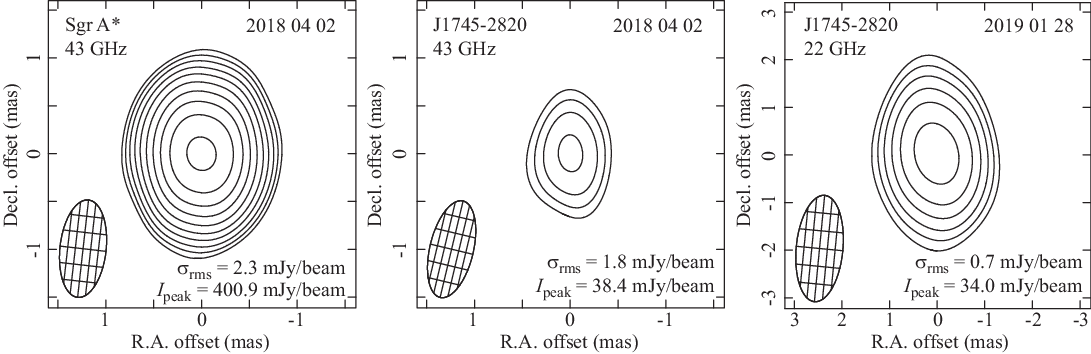} 
\end{center}
\caption{(Left) Synthesized image of Sgr A* at 43 GHz with the self-calibrated data. 
(Middle) Synthesized image of J1745$-$2820 at 43 GHz from the phase-referencing technique.
(Right) Synthesized image of J1745$-$2820 at 22 GHz with the self-calibrated data.
The peak intensity ($I_{\rm peak}$) and the rms noise level ($\sigma_{\rm rms}$)
are shown in the bottom right corner of each panel.
The contours are plotted at levels of 
$5\sigma_{\rm rms} \times \surd 2^n (n = 1, 2, 3, \cdots)$.}
\label{fig:5}
\end{figure*}

Figure \ref{fig:5} shows a synthesized image of Sgr A* using the self-calibrated data and that of J1745$-$2820 using the data calibrated in the phase-referencing technique at 43 GHz,
which were obtained from the 9th epoch data, taken with the newly developed broad-band recording system verified above, yielding the highest image dynamic range.
The peak intensity of Sgr A* during our monitoring observations was 300--700 mJy beam$^{-1}$; 
this average intensity variation was consistent with that reported by \citet{aki13}.
The structure of Sgr A* was very symmetric and well fitted to a single Gaussian component,
as reported in previous studies (\cite{lo98}; \cite{bow04}; \cite{she05}; \cite{lu11}; \cite{bow14}; \cite{joh18}; \cite{cho22}).
The peak intensity of J1745$-$2820 was 20--80 mJy beam$^{-1}$.
J1745$-$2820 was one order of magnitude weaker than Sgr A*, and
too week to map through by the self-calibration.
However, it can be mapped by phase-referencing using the calibration solution obtained from Sgr A*.
J1745$-$2820 on the phase-referenced image appeared to be point-like.
In addition, the self-calibration image of J1745$-$2820 at 22 GHz 
shown in figure \ref{fig:5} also appeared to be point-like.
Therefore, we simply obtained the position of J1745$-$2820 by a single Gaussian fitting.
Because the pair of Sgr A* and J1745$-$2820 is at low elevations, 
uncalibrated phase errors for the troposphere degraded the quality of the phase-referenced image.
Therefore, the phase-referenced image exhibits a low dynamic range of 7--28.

\begin{figure*}
\begin{center}
\includegraphics[width=17cm]{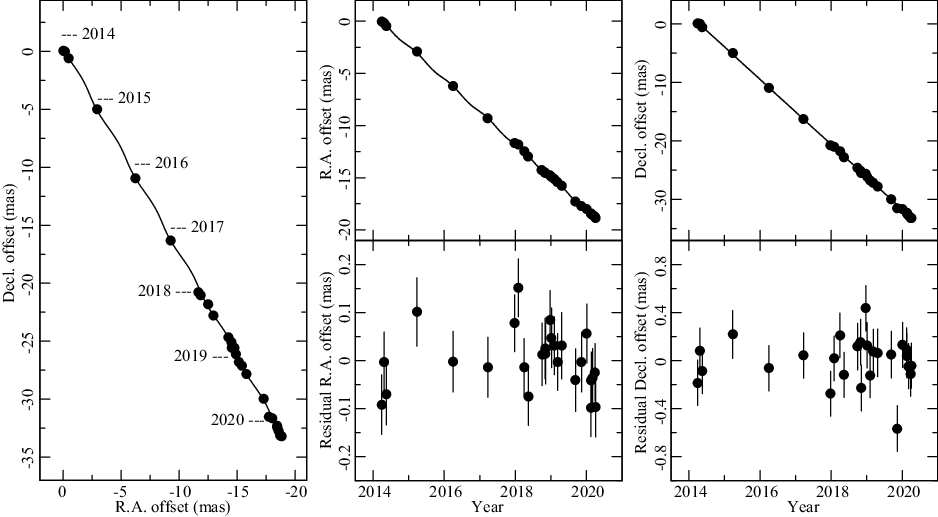} 
\end{center}
\caption{Proper motion of Sgr A*.
The solid lines show the modeled orbital motion,
which is sum of the best-fitted proper motion and the fixed parallax motion of 0.125 mas.
(Left) Positional variation on the sky. 
(Top-middle) RA position offset versus time.
(Bottom-middle) Same as the top-middle panel but with the best-fitted proper motion subtracted.
(Top-right) Same as the middle panels but for the Dec offset versus time.
(Bottom-right) Same as the top-right panel but with the best-fitted proper motion subtracted. }
\label{fig:6}
\end{figure*}

The measured positions of J1745$-$2820 with respect to Sgr A* 
at the individual epochs are summarized in table \ref{tab:2}.
From the inversion of these positions, 
we obtained the positional variation of Sgr A* from 2014 to 2020,
as shown in figure \ref{fig:6}.
We can clearly observe linear proper motion along the Galactic plane.
Using the Markov chain Monte Carlo method, 
we fit the measured positions of Sgr A*
with the position offset origin of $(\Delta \alpha_0, \Delta \delta_0)$
and the linear proper motion of $(\mu_\alpha, \mu_\delta)$.
At this time, the parallax is fixed to $\pi = 125$ $\mu$as, which corresponds to a distance of 8 kpc.
Therefore, the proper motion is obtained to be
$(\mu_\alpha, \mu_\delta) = (-3.136\pm0.007, -5.555\pm0.020)$ mas yr$^{-1}$ in RA and Dec.
The standard deviations of the post-fit residual were 
0.061 and 0.188 mas in RA and Dec, respectively.
The formal position errors ($\sigma_{\rm for}$) caused by thermal noise,
which are estimated for each observation epoch by using the 2D Gaussian fitting of 
the AIPS task IMFIT,
are in the ranges of 0.009--0.040 mas in RA and 0.020--0.073 mas in Dec.
The average of all epochs is 0.021 mas in RA and 0.040 mas in Dec.
The systematic position errors ($\sigma_{\rm sys}$), 
which are added in quadrature to the formal position error 
to achieve a reduced chi-square value of unity 
($\chi^2_\nu \approx 1$), 
are estimated to be 0.058 mas in RA and 0.184 mas in Dec.
The position errors are listed in table \ref{tab:2} and the error bars in figure \ref{fig:6}
represent the total position error estimated by
the root-square-sum (RSS) of formal and systematic position errors, 
$\surd(\sigma_{\rm for}^2 + \sigma_{\rm sys}^2)$.

\begin{figure*}
\begin{center}
\includegraphics[width=17cm]{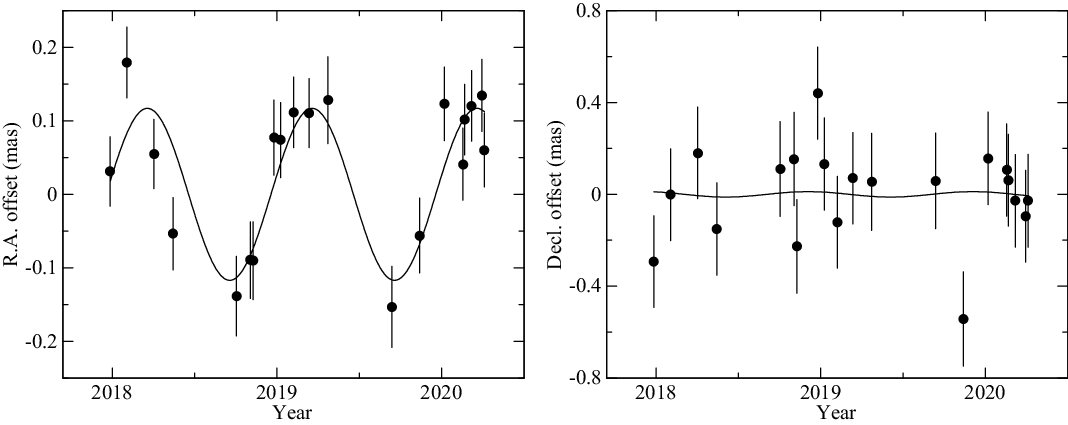} 
\end{center}
\caption{Trigonometric parallax of Sgr A*. 
The best-fit proper motion and the constant position offset are removed, 
allowing the effect of only the parallax to be seen.
The solid lines show the best-fitted parallax of $0.117\pm0.017$ mas, 
corresponding to a distance of $8.5^{+1.5}_{-1.1}$ kpc.
Position residuals in RA and Dec from the constant proper motion are shown in the left and right panels, respectively.}
\label{fig:7}
\end{figure*}

As shown in figure \ref{fig:7}, we can clearly observe sinusoidal parallax motion
with a period of one year in RA. 
This is the first detection of the parallax for Sgr A* using the data of 20 epochs 
from 7th to 26th as $\pi = 0.117\pm0.017$ mas (with a relative error of 15\%). 
For the sinusoidal parallax motion of Sgr A*, 
the amplitude of Dec was $\sim 1/10$ that of RA.
In addition, the astrometric position error in Dec was a factor of $\sim 2$--3 
larger than that in RA because of the larger synthesized beam size in Dec.
Therefore, no contribution from the Dec data was considered in the parallax measurements.
The obtained parallax corresponds to the Galactocentric distance of 
$R_0=8.5^{+1.5}_{-1.1}$ kpc.
This is consistent with the recently estimated Galactocentric distance of 
7.92--8.28 kpc (\cite{do19}; \cite{rei19}; \cite{vera20}; \cite{gra21}).
However, the present relative error of 15\% was an order of magnitude larger than that of the other measurements (0.3\%--4\%).
We continue the observation, and add $\approx 150$ epochs for the next five years.
The relative error would be statistically reduced to $15\% / \surd(170/20) \approx 5$\%.  

Hereafter, we assume that the distance from the LSR to Sgr A* and 
the Galactocentric distance of the LSR are 8 kpc.

\section{Discussion}

\subsection{Astrometric position error and core wander}

The post-fit residual regarded as the astrometric position error
would be affected by both calibration errors of the observations 
and the core position wander of Sgr A* which is caused by 
refractive interstellar scattering \citep{rei99, joh18} and gravitation of 
 intermediate-mass black holes (IMBHs) \citep{rei04, rei20}.
In this section, we investigate these effects quantitatively.

In VLBI astrometry, uncalibrated delays of the troposphere, 
ionosphere, station position, source position (and structure), 
and instrument propagate to the astrometric position error.
Table \ref{tab:4} summarizes the astrometric position error 
caused by each error source for Sgr A* observations.
They were estimated using the relationships,
\begin{eqnarray}
\Delta \theta \approx \frac{c \Delta \tau_{\rm err}}{D_{\rm proj}} 
              \approx \frac{c \Delta \tau_{\rm err}}{\lambda} \cdot \theta_{\rm beam},
\end{eqnarray}
where $c \Delta \tau_{\rm err}$ is the delay calibration error between 
the target and the reference source, 
$D_{\rm proj}$ is the projected baseline length, 
$\lambda = 7$ mm is the observing wavelength, and 
$\theta_{\rm beam}$ is the synthesized beam size, 
which is $\sim 0.5$ mas in RA and $\sim 1$ mas in Dec in our observations.

The delay calibration error, $c \Delta \tau_{\rm err}$, 
of each error source was estimated as follow.
$c \Delta \tau_{\rm err}$ of the troposphere is approximated to be 
$c\Delta\tau_{\rm err} \approx c\tau_{\rm err} \cdot \Delta \sec Z \approx 1.3$ mm
from the tropospheric zenith delay error of $c \tau_{\rm err} \approx 20$ mm \citep{hon08b} 
and the difference of $\sec Z$ between Sgr A* and J1745$-$2820 of 
$\Delta \sec Z \approx 0.065$ (section 3.2).
For the ionosphere, the error in the total electron content (TEC) 
is estimated to be $\sim 10$ TEC units (TECU) \citep{ho97},
which is equivalent to $c\tau_{\rm err} \approx 2$ mm at the observing frequency of 43 GHz.
The single-layer model mapping function is expressed as $\sec Z'$,
where $Z'$ is the source's zenith angle at the ionospheric pierce point given by
$\sin Z' = R/(R+H) \cdot \sin Z$ \citep{sch99}.
Here, $Z$ is the source's zenith angle at the station,
$R\approx6371$ km is the mean earth radius, and $H\approx450$ km is 
the height of the single-layer.
The difference in $\sec Z'$ between Sgr A* and J1745$-$2820 
is calculated to be $\Delta \sec Z'\approx 0.023$.
Therefore, 
$c \Delta \tau_{\rm err}$ of the ionosphere is estimated to be
$c\Delta\tau_{\rm err} \approx c\tau_{\rm err} \cdot \Delta \sec Z' \approx 0.04$ mm.
$c \Delta \tau_{\rm err}$ of the station position is also estimated to be
$c\Delta\tau_{\rm err} \approx c \tau_{\rm err} \cdot \theta_{\rm SA} \approx 0.04$ mm
from the station position error of $c\tau_{\rm err} \approx 3$ mm \citep{jik09, jik18} 
and the separation angle between Sgr A* and J1745$-$2820 
of $\theta_{\rm SA} \approx \timeform{0.67D} \approx 0.012$ rad.
The calibration error of the instrumental delay for the VERA dual-beam system
is independent of the target-reference-separation and is estimated to be 
$c \tau_{\rm err} \approx c \Delta \tau_{\rm err} \approx 0.1$ mm \citep{hon08a}.
The astrometric position errors caused 
by the source position error (section 3.2) and thermal noise (section 4.2) 
are listed in table \ref{tab:4}.
Comparison of the astrometric position errors shows that the troposphere dominated the other error sources.

As mentioned in section 4.2,
the standard deviations of the post-fit residuals were 0.061 mas in RA and 0.188 mas in Dec, respectively.
However, as shown in table \ref{tab:4}, 
the total position errors estimated from the RSS of the six error sources 
are 0.09 mas in RA and 0.20 mas in Dec.
These values are consistent within a factor of 1.5.

Refractive scattering can distort an image and cause position wandering of the core.
\citet{rei99} expect the refractive position wander of Sgr A* at 43 GHz
to be $< 0.04$ mas.
This is smaller than the estimated tropospheric error and the observed post-fit residual.

\begin{table}
\tbl{Theoretical estimation of astrometric position error for Sgr A* observations at 43 GHz.}{
\begin{tabular}{cccll}
\hline
\multicolumn{1}{c}{Error source}                &
\multicolumn{1}{c}{$c\tau_{\rm err}$}           &
\multicolumn{1}{c}{$c\Delta\tau_{\rm err}$}     &
\multicolumn{1}{c}{$\Delta \theta_\alpha$}      &
\multicolumn{1}{c}{$\Delta \theta_\delta$}      \\
                                                &
\multicolumn{1}{c}{(mm)}                        &
\multicolumn{1}{c}{(mm)}                        &
\multicolumn{1}{c}{(mas)}                   &
\multicolumn{1}{c}{(mas)}                   \\
\hline
troposphere        & 20  & 1.3  & 0.09  & 0.19  \\
ionosphere         & 2   & 0.04 & 0.003 & 0.005 \\
station coordinate & 3   & 0.04 & 0.003 & 0.005 \\
instrument         & 0.1 & 0.1  & 0.007 & 0.014 \\
source coordinate  & ... & ...  & 0.01  & 0.01  \\
thermal noise      & ... & ...  & 0.02  & 0.04  \\
\hline
Total (RSS)        &     &      & 0.09  & 0.20  \\
\hline
\end{tabular}}
\label{tab:4}
\begin{tabnote}
Column 1 lists  the sources of the errors. ``Total'' means the RSS of the six error sources.
Column 2 lists the excess path delay errors.
Column 3 lists the excess path delay errors between Sgr A* and J1745$-$2820.
Columns 4 and 5 list the astrometric position errors in RA and Dec, respectively. (Previous observations showed that Sgr A* is a point-like source (e.g., \cite{aki13}); therefore, the effect of its structure is negligible.)
\end{tabnote}
\end{table}

The core wander of Sgr A* caused by refractive interstellar scattering 
and gravitation of IMBHs 
is less than this level, with the upper limit of the core wander 
estimated to be $\surd(0.061^2+0.188^2) = 0.20$ mas (1.6 AU at 8 kpc).

\subsection{Peculiar motion and acceleration}

\begin{figure*}
\begin{center}
\includegraphics[width=17cm]{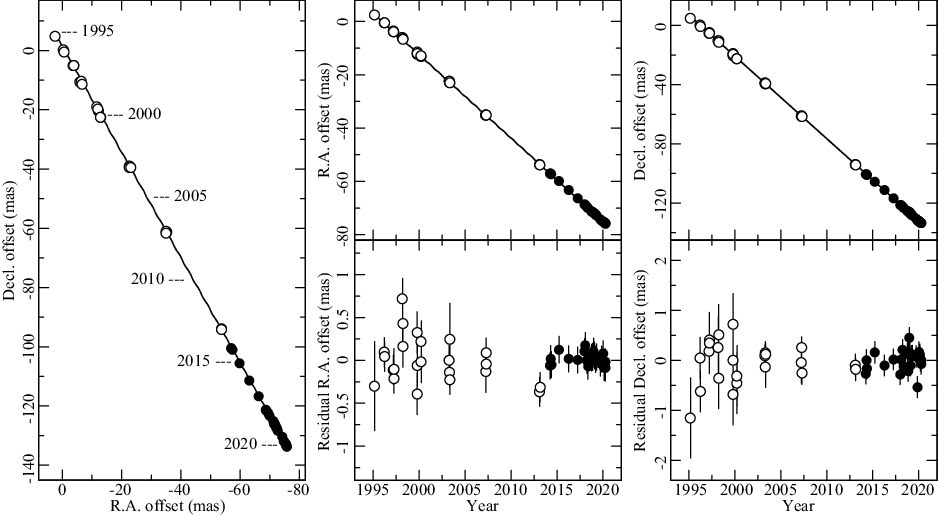} 
\end{center}
\caption{Location of Sgr A* 
combining the measurements of \citet{rei20} (open circles) and those of this work (filled circles).
The solid line shows the modeled motion of Sgr A*,
which is a sum of the best-fitted proper motion and the fixed parallax motion of 0.125 mas.
(Left) Position residuals on the sky. 
(Middle) RA offset versus time.
(Right) Same as the middle panels but for the Dec offset versus time.}
\label{fig:8}
\end{figure*}

The proper motion of Sgr A* with respect to J1745$-$2820 in RA and Dec
was measured in the period of 1995--2013 with the VLBA by \citet{rei20}, as 
\begin{eqnarray}
(\mu_\alpha, \mu_\delta) = (-3.147\pm0.008, -5.578\pm0.011)\ {\rm mas\ yr}^{-1},
\end{eqnarray}
and in the period of 2014--2020 with VERA in this paper was found to be,
\begin{eqnarray}
(\mu_\alpha, \mu_\delta) = (-3.136\pm0.007, -5.555\pm0.020)\ {\rm mas\ yr}^{-1}.
\end{eqnarray}
These results are consistent with each other.
The difference of 
$(\Delta \mu_\alpha, \Delta \mu_\delta) = (0.011\pm0.011, 0.023\pm0.023)$ mas yr$^{-1}$
 is not significant.
We obtained better estimate of the proper motion by combining two measurements.
To compensate for differences in the source positions of the correlation for each reference epoch (1st epoch), we 
corrected the source position differences between those of \citet{rei20} and us:
$(\Delta \alpha, \Delta \delta) = (70.98, 95.68)$ mas for Sgr A* 
and $(14.04, -4.85)$ mas for J1745$-$2820. 
Position offsets based on the source position of \citet{rei20} 
are listed in columns 6 and 7 of table \ref{tab:2}.
Figure \ref{fig:8} shows the proper motion of Sgr A* obtained from a combination of the astrometric results of \citet{rei20} and us.
Therefore the proper motion in RA and Dec is obtained as
\begin{eqnarray}
(\mu_\alpha, \mu_\delta) = (-3.133\pm0.003, -5.575\pm0.005)\ {\rm mas\ yr}^{-1}.
\end{eqnarray}
The standard deviations of post-fit residuals
for \citet{rei20}, our data, and all of them are,
(0.26, 0.41), (0.06, 0.19), and (0.19, 0.32) mas, respectively, in RA and Dec.
The systematic error of (0.14, 0.08) mas is added in quadrature to 
the position errors of both \citet{rei20} and our data 
to achieve a $\chi^2_\nu \approx 1$.
This result is consistent with the latest updated proper motion using the VLBA by \citet{xu22}.

Note that there is an uncertainty of $\approx 0.1$ mas 
in the position comparison between \citet{rei20} and us for Sgr A*.
Because the source position of J1745$-$2820 is inaccurate by $\approx 10$ mas (section 3.2),
its position adopted in the delay tracking differs by 
$(\Delta \alpha, \Delta \delta) = (14.04, -4.85)$ mas between the two measurements.
This difference for J1745$-$2820 could cause a systematic offset in the position comparison for Sgr A*. 
The offset can be roughly estimated using the separation angle of 
$\theta_{\rm SA}\approx\timeform{0.67D}\approx0.012$ rad to be 
$(|\Delta \alpha \cdot \theta_{\rm SA}|, |\Delta \delta \cdot \theta_{\rm SA}|)\approx(0.17, 0.06)$ mas,
and is close to the systematic error of (0.14, 0.08) mas.

The proper motion in RA and Dec in equation (4) can thus be converted to 
\begin{eqnarray}
(\mu_l, \mu_b) = (-6.391\pm0.005, -0.230\pm0.004)\ {\rm mas\ yr}^{-1}
\end{eqnarray}
in Galactic longitude and latitude.
The motion in the Galactic longitude direction, $\mu_l = -6.391\pm0.005$ mas yr$^{-1}$,
corresponds to the angular velocity of the Sun,
which is the sum of the angular velocities of the Galactic rotation at the LSR and 
the solar motion in the direction of Galactic rotation,
$\Omega_{\odot} = -4.74074 \cdot \mu_l = 30.30\pm0.02$ km s$^{-1}$ kpc$^{-1}$.
When we adopt $R_0 = 8$ kpc and solar motion toward $l = \timeform{90D}$
of $V_{\odot} = 12.2\pm2.5$ km s$^{-1}$ \citep{sch10},
the angular velocity of the Galactic rotation at the LSR 
is estimated to be $\Omega_0 = \Omega_\odot - V_\odot / R_0 = 28.77\pm0.31$ km s$^{-1}$ kpc$^{-1}$.
This is 11\% larger than 
the International Astronomical Union (IAU) recommended value of 
(220 km s$^{-1}$)/(8.5 kpc) = 25.9 km s$^{-1}$ kpc$^{-1}$ \citep{ker86}.
At $R_0=8$ kpc, the Galactic rotation velocity of the LSR is derived to be 
$\Theta_0=\Omega_0 \cdot R_0 = 230\pm3$ km s$^{-1}$.
The proper motion in the Galactic latitude direction, 
$\mu_b = -0.230\pm0.004$ mas yr$^{-1}$ or $-8.7\pm0.2$ km s$^{-1}$ at 8 kpc,
is consistent with the inverse of the solar motion toward the north Galactic pole of 
$W_{\odot} = 7.3\pm0.9$ km s$^{-1}$ \citep{sch10}.

After subtracting the Galactic rotation at the LSR of 
$\Omega_{\odot} = 30.17\pm0.40$ km s$^{-1}$ kpc$^{-1}$, which is  
estimated on the basis of the analysis of VLBI astrometric results of 189 maser sources \citep{vera20},
and $W_{\odot} = 7.3\pm0.9$ km s$^{-1}$ \citep{sch10},
the peculiar motion of Sgr A* with respect to 
the dynamical center of the Galactic rotation was estimated to be
$(\Delta \mu_l, \Delta \mu_b) = (-0.027\pm0.085,-0.038\pm0.024)$ mas yr$^{-1}$
or $(\Delta v_l, \Delta v_b) = (-1.0\pm3.2, -1.4\pm0.9)$ km s$^{-1}$ at 8 kpc.
The upper limit was estimated to be
$\surd [(-0.027)^2+(0.085)^2+(-0.038)^2+(0.024)^2] = 0.10$ mas yr$^{-1}$ (3.7 km s$^{-1}$ at 8 kpc).

As shown in equations (2) and (3), by comparing the proper motion as measured by \citet{rei20} to ours,
it is clear that the proper motion of Sgr A* has not changed significantly over the past few decades.
If we fit the positions of the combined data with the two measurements using a second-order polynomial
(i.e., $\Delta \alpha=\pi_\alpha(t)+\Delta \alpha_0+\mu_\alpha t + 0.5 a_\alpha t^2$), 
the acceleration in RA and Dec is obtained as
\begin{eqnarray}
(a_\alpha,a_\delta) = (1.5\pm1.2,0.1\pm1.8) \ \mu{\rm as\ yr}^{-2}.
\end{eqnarray}
This acceleration value is consistent with zero within $1\sigma$.
The upper limit was estimated to be
$\surd [(1.5)^2+(1.2)^2+(0.1)^2+(1.8)^2] = 2.6$ $\mu$as yr$^{-2}$ 
(0.10 km s$^{-1}$ yr$^{-1}$ at 8 kpc). 

\subsection{Limits on the mass of intermediate-mass black hole (IMBH)}
In the Galactic center region, a nuclear star cluster (NSC) is present despite the existence of strong tidal forces. 
However, its formation process is still under discussion. Several scenarios regarding in situ or outside formation processes at the central parsec as the origin of the young stars have been proposed \citep{ger01, ghe03, gou03, lev07, yel14, fel14, sch20}. 
For the outside formation scenario, there are some mechanisms proposed to move star clusters to the Galactic center. 
One is that clusters accompanied by IMBHs fall into the Galactic center through dynamical friction \citep{ebi01, fuj09, arc18}.
IMBHs may possibly merge to form an SMBH and/or move around Sgr A*.
Within the last 20 years, many IMBH candidates have been discovered in the Galactic center region \citep{mai04, oka16, tub17, tak20}.
In particular, one of the candidates, IRS13E, is located as close as 0.1 pc to the Galactic center, and whether IRS13E harbors an IMBH is currently under debate \citep{sch05, pau06, fri10, tub20, zhu20}.

The possibility of the existence of IMBHs near Sgr A* has been investigated by finding the orbital perturbations caused by their gravitational interactions with Sgr A* or S2, which is the star closest to Sgr A*, and the gravitational radiation as the bremsstrahlung of IMBHs resulting from dynamical effects \citep{han03, yu03, gua09, gir19, nao20, fra20, gra23}. 
In this paper, we mainly focus on the reflex motion of Sgr A* due to an IMBH perturbation, such as the acceleration $a$, peculiar motion $\Delta v$, and core wander $\Delta \theta$ 
, whose limits are given from the limitation of astrometric accuracy described in sections 5.1 and 5.2. 
The limitations to the mass and the location of an IMBH from $\Delta \theta$, $\Delta v$ and $a$ can be calculated using the following equations:  

\begin{eqnarray}
\label{equ:7}
r \lesssim \frac{\Delta \theta (M_{\rm Sgr A^*} +M)}{M},
\end{eqnarray}
\begin{eqnarray} 
\label{equ:8}
r \gtrsim \frac{GM^2}{(\Delta v)^2 (M_{\rm Sgr A^*} +M)},
\end{eqnarray}
\begin{eqnarray}
\label{equ:9}
a = \frac{GM}{r^2}, 
\end{eqnarray} 
where $G$ is the gravitational constant, $M$ is the mass of an IMBH, and $r$ is the orbital radius. Equations (7) and (8) are cited from equations (36) and (37) of \citet{yu03}, respectively. The mass of $M_{\rm Sgr A^*} =$ 4 $\times$ 10$^6$$\MO$ is adopted for Sgr A* here (e.g., 4.297 $\times$ 10$^6$$\MO$, \cite{gra21}). 

\begin{figure}
\begin{center}
\includegraphics[width=8cm]{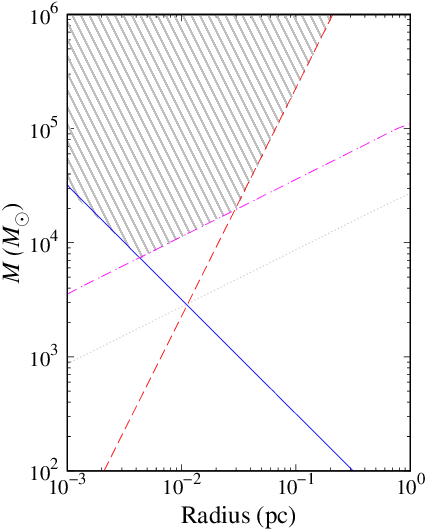} 
\end{center}
\caption{Limits on the mass of an IMBH. Its horizontal and vertical axes show the orbital radius and the mass of an IMBH, respectively. The blue line shows the limit by astrometric accuracy derived using $\Delta\theta < 0.20$ mas. The dashed red line is derived using acceleration ($a < 0.10$ km s$^{-1}$ yr$^{-1}$). The dotted-dashed magenta line is derived using $\Delta v < 3.7$ km s$^{-1}$. 
The dotted gray line ($\Delta v = 0.9$ km s$^{-1}$) indicates the position where the three lines intersect.
The IMBH is excluded to exist in the shaded area by all three constraints.
}
\label{fig:9}
\end{figure}

The estimated limits on the parameters are shown in figure \ref{fig:9}. 
The upper limits to the IMBH mass are $\approx$ 3 $\times$ 10$^{4}$$M_{\odot}$ and  $\approx$ 3 $\times$ 10$^{3}$$M_{\odot}$ at 0.1 and 0.01 pc from the Galactic center, respectively. 
By comparing our results with those of \citet{rei20}, the excluded parameter region can be extended for the following reasons; (1) the addition of our data of dense observations for seven years to Reid's pioneering long-term observations (18 yr) limits the acceleration from 0.38 to 0.1 km s$^{-1}$, and (2) dual-beam, high-sensitivity broad-band observations improve the accuracy of the $\Delta\theta$ measurement from 0.5 to 0.2 mas.
This result is an update of and is also well consistent with those of \citet{yu03, gua09, rei20}. 

\section{Conclusion}

We have conducted the VLBI astrometric observations of Sgr A* at 43 GHz with VERA.
The present conclusions are summarized as follows:

\begin{enumerate}
  \item We confirmed that the three data acquisition (sampling) systems of ADS1000, ADS3000+ and OCTAD are consistent 
  at the 10-$\mu$as level in astrometric position and at the 10\% level in the amplitude and SNR.
  \item Both the target Sgr A* and the position reference source J1745$-$2820 
  could be detected in the longest baselines of VERA using the newly developed broad-band receiving system with 
  a sampling rate of 8 Gbps, providing a total bandwidth of 4 GHz.
  \item The trigonometric parallax of Sgr A* was obtained to be $0.117 \pm 0.017$ mas,
  corresponding to a Galactocentric distance of $R_0=8.5^{+1.5}_{-1.1}$ kpc. 
  This is the first parallax measurement for Sgr A*.
  \item The proper motion of Sgr A* measured in this paper and \citet{rei20} 
  are consistent at the 1\% level.
  Combining two measurements yields a proper motion of 
  $(\mu_\alpha, \mu_\delta) = (-3.133\pm0.003, -5.575\pm0.005)$ mas yr$^{-1}$
  in RA and Dec and 
  $(\mu_l, \mu_b) = (-6.391\pm0.005, -0.230\pm0.004)$ mas yr$^{-1}$ 
  in the Galactic longitude and latitude.
  From the proper motion in the Galactic longitudinal direction,
  the angular velocity of the Sun is estimated to be
  $\Omega_\odot = 30.30 \pm 0.02$ km s$^{-1}$ kpc$^{-1}$.
  \item The 1$\sigma$ upper limit of the core wander ($\Delta \theta$), 
  peculiar motion ($\Delta v$), and acceleration ($a$) of Sgr A* are
  estimated to be $\Delta \theta < 0.20$ mas (1.6 AU), 
  $\Delta v < 0.10$ mas yr$^{-1}$ (3.7 km s$^{-1}$), and 
  $a < 2.6$ $\mu$as yr$^{-2}$ (0.10 km s$^{-1}$ yr$^{-1}$), respectively.
  Therefore, Sgr A* appears to be at rest with respect to the dynamical center of the Galactic rotation.
  \item The upper mass limits on the IMBH masses are  $\approx$ 3 $\times$ 10$^{4}$$M_{\odot}$ and  $\approx$ 3 $\times$ 10$^{3}$$M_{\odot}$ at 0.1 and 0.01 pc from the Galactic center, respectively
\end{enumerate}

\begin{ack}
Data analysis was (in part) performed on the Multi-wavelength Data Analysis System 
operated by the Astronomy Data Center of the National Astronomical Observatory of Japan.
This work was partially supported by JSPS KAKENHI Grant Numbers JP24684011 (TH), JP17K05398 (TH), JP15H03644 (YH and YK), JP17H01116 (MH), JP18KK0090 (KH), JP19H01943 (KH and YH), and JP22H00157 (KH). The development of the new VERA signal transmission system was partially supported by the Amanogawa Galaxy Astronomy Research Center at Kagoshima University. 
\end{ack}







\begin{thebibliography}{}
\bibitem[Akiyama et al.(2013)]{aki13}
Akiyama,~K., Takahashi,~R., Honma,~M., Oyama,~T., \& Kobayashi,~H. 2013, \pasj, 65, 91
\bibitem[Arca-Sedda \& Gualandris(2018)]{arc18}
Arca-Sedda,~M., \& Gualandris,~A. 2018, \mnras, 477, 4423
\bibitem[An et al.(2018)]{an18}
An,~T., Sohn, B.~W., \& Imai,~H. 2018, Nature Astron., 2, 118
\bibitem[Backer \& Sramek(1999)]{bac99}
Backer, D.~C., \& Sramek, R.~A. 1999, \apj, 524, 805
\bibitem[Bower et al.(2004)]{bow04}
Bower, G.~C, Falcke,~H., Herrnstein, R.~M., Zhao, J.-H., Goss, W.~M., \& Backer, D.~C. 2004, Science, 304, 704
\bibitem[Bower et al.(2014)]{bow14}
Bower, G.~C, \etal\ 2014, \apj, 790, 1
\bibitem[Cho et al.(2022)]{cho22}
Cho,~I. \etal\ 2022, \apj, 926, 108
\bibitem[Do et al.(2019)]{do19}
Do,~T., \etal\ 2019, Science, 365, 664
\bibitem[Doi et al.(2009)]{doi09}
Doi,~A., \etal\ 2009, \pasj, 61, 1389
\bibitem[Ebisuzaki et al.(2001)]{ebi01}
Ebisuzaki,~T., \etal\ 2001, \apj, 562, L19
\bibitem[Event Horizon Telescope Collaboration et al.(2022)]{eht22}
Event Horizon Telescope Collaboration, \etal\ 2022, \apjl, 930, L12
\bibitem[Feldmeier et al.(2014)]{fel14}
Feldmeier,~A., Neumayer,~N., Seth,~A., \etal\  2014, \aap, 570, A2
\bibitem[Fragione et al.(2020)]{fra20}
Fragione,~G., Loeb,~A., Kremer,~K., \& Rasio,  F.~A. 2020, \apj, 897, 46
\bibitem[Fritz et al.(2010)]{fri10}
Fritz, T.~K., Gillessen,~S., Dodds-Eden,~K., \etal\ 2010, \apj, 721, 395
\bibitem[Fujii et al.(2009)]{fuj09}
Fujii,~M., Iwasawa,~M., Funato,~Y., \& Makino,~J. 2009, \apj,  695, 1421
\bibitem[Fujisawa et al.(2001)]{fuj01}
Fujisawa,~K., \etal\ 2001, Communications Research Laboratory Journal, 48, 47
\bibitem[Gerhard(2001)]{ger01}
Gerhard,~O. 2001, \apjl, 546, L39
\bibitem[Ghez et al.(2003)]{ghe03}
Ghez, A.~M., \etal\ 2003, \apjl, 586, L127
\bibitem[Girma \& Loeb(2019)]{gir19}
Girma,~E., \& Loeb,~A. 2019, \mnras, 482, 3669
\bibitem[Gould \& Quillen(2003)]{gou03}
Gould,~A., \& Quillen, A.~C. 2003, \apj, 592, 935
\bibitem[GRAVITY Collaboration et al.(2019)]{gra19}
GRAVITY Collaboration, \etal\ 2019, \aap, 625, L10
\bibitem[GRAVITY Collaboration et al.(2021)]{gra21}
GRAVITY Collaboration, \etal\ 2021, \aap, 647, A59
\bibitem[GRAVITY Collaboration et al.(2023)]{gra23}
GRAVITY Collaboration, \etal\ 2023, \aap, 672, A63
\bibitem[Gualandris \& Merritt(2009)]{gua09}
Gualandris,~A., \& Merritt,~D. 2009, \apj, 705, 361
\bibitem[Gwinn et al.(2014)]{gwi14}
Gwinn, C.~R., Kovalev, Y.~Y., Johnson, M.~D., \& Soglasnov, V.~A. 2014, \apjl, 794, L14
\bibitem[Hagiwara et al.(2022)]{hag22}
Hagiwara,~Y., Hada,~K., Takamura,~M., Oyama,~T., Yamauchi,~ A., \& Suzuki,~S. 2022, Galaxies, 10, 114
\bibitem[Hansen \& Milosavljevi{\'c}(2003)]{han03}
Hansen, B.~M.~S., \& Milosavljevi{\'c}, M. 2003, \apjl, 593, L77
\bibitem[Hasegawa et al.(2004)]{has04}
Hasegawa,~T., Hasegawa,~T., Kawaguchi,~N., Fujisawa,~K., Takashima,~K., Uose,~H., 
\& Asano,~S. 2004, IEICE Trans. Commun., E87-B, 651
\bibitem[Ho et al.(1997)]{ho97} 
Ho, C.~M., Wilson, B.~D., Mannucci, A.~J., Lindqwister, U.~J., \& Yuan, D.~N. 1997, Radio Sci., 32, 1499
\bibitem[Honma et al.(2008a)]{hon08a}
Honma,~M., \etal\ 2008a, \pasj, 60, 935 
\bibitem[Honma et al.(2008b)]{hon08b}
Honma,~M., Tamura,~Y., \& Reid, M.~J. 2008b, \pasj, 60, 951
\bibitem[Iguchi et al.(2005)]{igu05}
Iguchi,~S., Kurayama,~T., Kawaguchi,~N., \& Kawakami,~K. 2005, \pasj, 57, 259
\bibitem[Jike et al.(2009)]{jik09} 
Jike,~T., Manabe,~S., \& Tamura,~Y. 2009, Journal of the Geodetic Society of Japan, 55, 369
\bibitem[Jike et al.(2018)]{jik18}
Jike,~T., Manabe,~S., \& Tamura,~Y. 2018, Journal of the Geodetic Society of Japan, 63, 193
\bibitem[Johnson et al.(2018)]{joh18}
Johnson, M.~D., \etal\ 2018, \apj, 865, 104
\bibitem[Kameno et al.(2014)]{kam14}
Kameno,~S., \etal\ 2014, \procspie, 9153, 91532D
\bibitem[Kawaguchi et al.(2000)]{kaw00} 
Kawaguchi,~N., Sasao,~T., \& Manabe,~S. 2000, \procspie, 4015, 544
\bibitem[Kawaguchi et al.(2001)]{kaw01}
Kawaguchi,~N., Fujisawa,~K., Nakajima,~J., Uose,~H., Iwamura,~S., Hoshino,~T., 
Hashimoto,~T., \& Takagi,~H. 2001, NTT R\&D, 50, 824
\bibitem[Kerr \& Lynden-Bell(1986)]{ker86}
Kerr, F.~J., \& Lynden-Bell,~D. 1986, \mnras, 221, 1023
\bibitem[Kimura \& Nakajima(2002)]{kim02}
Kimura,~M., \& Nakajima,~J. 2002, IVS CRL Tech. Development Center News, 21, 31
\bibitem[Kobayashi et al.(2003)]{kob03}
Kobayashi,~H., \etal\ 2003, in ASP Conf. Ser., 306, New Tech-nologies in VLBI, ed. Y.~C. Minh (San Francisco, CA: ASP), 367
\bibitem[Kono et al.(2012)]{kon12}
Kono,~Y., \etal\ 2012, in Proc. Seventh General Meeting (GM2012) of the International VLBI Service for Geodesy and Astrometry (IVS), ed. D.~Behrend \& K.~D. Baver (Washington, D.C.: NASA), 96
\bibitem[Levin(2007)]{lev07}
Levin,~Y. 2007, \mnras, 374, 515
\bibitem[Lo et al.(1998)]{lo98}
Lo, K.~Y., Shen, Z.-Q., Zhao, J.-H., \& Ho, P.~T.~P., 1998, \apjl, 508, L61
\bibitem[Lu et al.(2011)]{lu11}
Lu, R.-S., Krichbaum, T.~P., Eckart,~A., K\"{o}nig,~S., Kunneriath,~D., Witzel,~G., Witzel,~A. \& Zensus, J.~A. 2011, \aap, 525, A76
\bibitem[Maillard et al.(2004)]{mai04}
Maillard, J.~P., Paumard,~T., Stolovy, S.~R., \& Rigaut,~F. 2004, \aap, 423, 155
\bibitem[Mizuno et al.(2014)]{miz14}
Mizuno,~I., \etal\ 2014, Journal of Astronomical Instrumentation, 3, 1450010
\bibitem[Nagayama et al.(2020)]{nag20}
Nagayama,~T., \etal\ 2020, \pasj, 72, 52
\bibitem[Naoz et al.(2020)]{nao20}
Naoz,~S., Will, C.~M., Ramirez-Ruiz,~E., Hees, A., Ghez, A.~M., \& Do,~T. 2020, \apjl, 888, L8
\bibitem[Niell(1996)]{nie96}
Niell, A.~E. 1996, J. Geophys. Res., 101, 3227
\bibitem[Oka et al.(2016)]{oka16}
Oka,~T., Mizuno,~R., Miura,~K., \& Takekawa,~S. 2016, \apjl, 816, L7
\bibitem[Oyama et al.(2012)]{oya12}
Oyama,~T., \etal\ 2012, in Proc. Seventh General Meeting (GM2012) of the International VLBI Service for Geodesy and Astrometry (IVS), ed. D.~Behrend \& K.~D. Baver (Washington, D.C.: NASA), 91
\bibitem[Oyama et al.(2016)]{oya16}
Oyama,~T., \etal\ 2016, \pasj, 68, 105
\bibitem[Paumard et al.(2006)]{pau06}
Paumard,~T., Genzel,~R., Martins,~F., \etal\ 2006, \apj, 643, 1011
\bibitem[Petrov et al.(2007)]{pet07}
Petrov,~L., Hirota,~T., Honma,~M., Shibata, K.~M., Jike,~T., \& Kobayashi,~H. 2007, \aj, 133, 2487
\bibitem[Reid et al.(1988)]{rei88}
Reid, M.~J., Schneps, M.~H., Moran, J.~M., Gwinn, C.~R., Genzel,~R., Downes,~D., \& Roennaeng,~B. 1988, \apj, 330, 809
\bibitem[Reid et al.(1999)]{rei99}
Reid, M.~J., Readhead, A.~C.~S., Vermeulen, R.~C., \& Treuhaft, R.~N. 1999, \apj, 524, 816
\bibitem[Reid \& Brunthaler(2004)]{rei04}
Reid, M.~J., \& Brunthaler,~A. 2004, \apj, 616, 872
\bibitem[Reid et al.(2009)]{rei09}
Reid, M.~J., Menten, K.~M., Zheng, X.~W., Brunthaler,~A., \& Xu,~Y. 2009, \apj, 705, 1548
\bibitem[Reid et al.(2019)]{rei19}
Reid, M.~J., et al.\ 2019, \apj, 885, 131
\bibitem[Reid \& Brunthaler(2020)]{rei20}
Reid, M.~J., \& Brunthaler, A. 2020, \apj, 892, 39
\bibitem[Sakai et al.(2017)]{sak17}
Sakai,~D., Oyama,~T., Nagayama,~T., Honma,~M., \& Kobayashi,~H. 2017, \pasj, 69, 64
\bibitem[Sakai et al.(2023)]{sak23}
Sakai,~D., Oyama,~T., Nagayama,~T., Honma,~M., \& Kobayashi,~H. 2023, \pasj, 75, 937
\bibitem[Schaer(1999)]{sch99}
Schaer,~S.,\ 1999, 
Mapping and Predicting the Earth's Ionosphere Using the Global Positioning System,
Geod. Geophys. Arb. Schweiz., vol. 59, Inst. fur Geod. und Photogramm., Zurich, Switzerland.
\bibitem[Sch{\"o}nrich et al.(2005)]{sch05}
Sch{\"o}nrich,~R., Eckart,~A., Iserlohe,~C., Genzel,~R., \& Ott,~T. 2005, \apj, 625, 111
\bibitem[Sch{\"o}nrich et al.(2010)]{sch10}
Sch{\"o}nrich,~R., Binney,~J., \& Dehnen,~W. 2010, \mnras, 403, 1829
\bibitem[Sch{\"o}del et al.(2020)]{sch20}
Sch{\"o}del,~R., Nogueras-Lara,~F., Gallego-Cano,~E., Shahzamanian,~B., Gallego-Calvente, A.~T., \& Gardini,~A. 2020, \aap, 641, A102
\bibitem[Shen et al.(2005)]{she05}
Shen, Z.-Q., Lo, K.~Y., Liang, M.-C., Ho, P.~T.~P., \& Zhao, J.-H. 2005, \nat, 438, 62
\bibitem[Takaba et al.(2008)]{tak08}
Takaba,~H., \etal\ 2008, Journal of the Geodetic Society of Japan, 54, 269
\bibitem[Takamura et al.(2023)]{tak23}
Takamura,~M. \etal\ 2023, \apj, 952, 47
\bibitem[Takefuji et al.(2010)]{tak10}
Takefuji,~K., Takeuchi,~H., Tsutsumi,~M., \& Koyama,~Y. 2010, Sixth International VLBI Service for Geodesy and Astronomy. Proceedings from the 2010 General Meeting, 378
\bibitem[Takekawa et al.(2020)]{tak20}
Takekawa,~S., Oka,~T., Iwata,~Y., Tsujimoto,~S., \& Nomura,~M. 2020, \apj, 890, 167
\bibitem[Takeuchi et al.(2006)]{tak06} 
Takeuchi,~H., Kimura,~M., Nakajima,~J., Kondo,~T., Koyama,~Y., Ichikawa,~R., Sekido,~M., \& Kawai,~E. 2006, \pasp, 118, 1739
\bibitem[Tsuboi et al.(2017)]{tub17}
Tsuboi,~M., Kitamura,~Y., Tsutsumi,~T., Uehara,~K., Miyoshi,~M., Miyawaki,~R., \& Miyazaki,~A. 2017, \apjl, 850, L5
\bibitem[Tsuboi et al.(2020)]{tub20}
Tsuboi,~M., Kitamura,~Y., Tsutsumi,~T., Miyawaki,~R., Miyoshi,~M., \& Miyazaki,~A. 2020, \pasj, 72, L5
\bibitem[VERA Collaboration et al.(2020)]{vera20} 
VERA Collaboration, \etal\ 2020, \pasj, 72, 50
\bibitem[Xu et al.(2022)]{xu22} 
Xu,~S., Zhang,~B., Reid, M.~J., Zheng,~X., Wang,~G., \& Jung,~T. 2022, \apj, 940, 15
\bibitem[Yelda et al.(2014)]{yel14}
Yelda,~S., Ghez, A.~M., Lu, J.~R., Do,~T., Meyer,~L., Morris, M.~R., \& Mathews,~K. 2014, \apj, 783, 131
\bibitem[Yu \& Tremaine(2003)]{yu03}
Yu,~Q., \& Tremaine,~S. 2003, \apj, 599, 1129
\bibitem[Zhu et al.(2020)]{zhu20}
Zhu,~Z., Li,~Z., Ciurlo,~F., Morris, M.~R., Zhang,~M., Do,~T., \& Ghez, A.~M. 2020, \apj, 897, 135
\end{thebibliography}
\end{document}